\documentclass{article}
\usepackage[T2A]{fontenc}
\usepackage[utf8]{inputenc}
\usepackage[english]{babel}
\usepackage[dvips]{graphicx}
\usepackage{caption}
\usepackage{float}
\graphicspath{{images/}}
\usepackage[toc,title,page]{appendix}
\usepackage{verbatim}
\usepackage{multirow}

\usepackage[hcentering, bindingoffset = 10mm, right = 15 mm, left = 10 mm, top=20 mm, bottom = 20 mm]{geometry} 

\DeclareGraphicsExtensions{.HEIC,.pdf,.png,.jpg} 
\usepackage{mathtext,cite, tikz, amsmath, amsfonts, amssymb, amsthm, mathtools, multirow, lipsum, amsmath, amstext, subcaption, upgreek, wrapfig, adjustbox, icomma, amssymb, enumerate, indentfirst, float, mathrsfs, hyperref, gensymb, chngpage, pgfplots, hhline, tabularx, xfrac, tikz-feynman, physics}
\usepackage{physics}

\title{Does the existence of $f_0$  meson enlarge the $K^+ \to \pi^0\pi^0\pi^0 e^+ \nu$ decay probability?}
\author{ \bfseries E. K. Karkaryan   \footnote{karkaryan@bk.ru}\\
 \textit{I.E. Tamm Department of Theoretical Physics, Lebedev Physical Institute,}\\
 \textit{53 Leninskiy Prospekt, Moscow, 119991, Russia} \\
 [2ex]
\bfseries V. F. Obraztsov\\
 \textit{NRC "Kurchatov Institute" -IHEP,}\\
 \textit{142281 Protvino, Russia}\\[2ex]
\bfseries M. I. Vysotsky\\
\textit{I.E. Tamm Department of Theoretical Physics, Lebedev Physical Institute,}\\
 \textit{53 Leninskiy Prospekt, Moscow, 119991, Russia} \\
 [2ex]}
\date{}

\begin{document}

\maketitle

\begin{abstract}
We estimate a possible contribution of the two-pion system $f_0(500)$ wide resonance to the decay probability of the super rare decay $K^+ \to 3\pi^0 e^+ \nu$. We vary the width of the resonance from 0.01 MeV to 1000 MeV. We found that the contribution of the narrow resonance in a system of several final state particles into the decay probability is proportional to its width. However it never exceeds the value of the decay probability calculated on free final state particles neglecting the resonance contribution.
\end{abstract}

The OKA collaboration has improved the upper limit of the $K^+ \to 3\pi^0 e^+ \nu$ decay probability. The new experimental result is: $Br\qty(K^+ \to 3\pi^0 e^+ \nu) < 5.4 \cdot 10^{-8}$ at the $90\%$ confidence level \cite{oka}. It is by more than four orders of magnitude larger than the chiral perturbation theory estimate: $Br\qty(K^+ \to 3\pi^0 e^+ \nu) \approx 2.5 \cdot 10^{-12}$  \cite{blaser}. It is evident that in order to detect this decay it's probability should be strongly enhanced. As it was noticed in \cite{blaser} the suppression of the $K^+ \to 3\pi^0 e^+ \nu$ decay probability by more than six orders of magnitude in comparison  with that of the $K^+ \to 2\pi^0 e^+ \nu$ decay  is almost entirely due to the reduction of the phase space. That is why in \cite{oka} it was suggested that pionium production in the  $K^+ \to A_{2\pi}\pi^0 e^+ \nu$ decay followed by $A_{2\pi} \to \pi^0\pi^0$ decay may considerably enhance $K^+ \to 3\pi^0 e^+ \nu$ decay probability as the five-particles final state phase volume is replaced by the four-particles one. As it was found in our recent paper \cite{pionium} the enhancement  is really huge and the decay through pionium is by almost five orders of magnitude more probable than the direct decay from the phase space point of view (see Eq.(13) in \cite{pionium}). However, the probability of the pionium production is proportional to its width. Pionium is $\pi^+\pi^-$ atom loosely bound by the Coulomb interaction. The width of the $A_{2\pi} \to \pi^0\pi^0$ decay equals $\sim 10^{-7}~\text{MeV}$ making the probability of the $K^+ \to A_{2\pi}\pi^0 e^+ \nu$ decay completely negligible. However there exists a two pion system resonance $f_0$  whose width approximately equals its mass: $\Gamma_{f_0} \approx M_{f_0}\approx 500~\text{MeV}$ according to RPP last edition \cite{rpp}. So, may be $f_0$ can do the job and the chain $K^+ \to f_0 \pi^0 e^+\nu \to \pi^0 \pi^0 \pi^0 e^+ \nu$ is enhanced in comparison with the direct decay?

In order to take into account the contribution of $f_0$ meson it is convenient to present the five particles final state in a form shown in Fig.\ref{1}, separating the contribution of the two $\pi^0$-mesons.

\begin{figure}[h]  \label{fig1}
    \centering
    \includegraphics[scale=1.5]{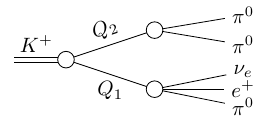}
    \caption{Convenient separation of final particles}
    \label{fig:enter-label}
\end{figure}

In this way neglecting an electron mass for the phase volume we get:

\begin{equation}\label{1}
V_5 = \int \frac{dQ^2_2}{2\pi} \int \frac{dQ^2_1}{2\pi} V_2(m^2_K; Q^2_1, Q^2_2)V_3\qty(Q^2_1)V_2(Q^2_2),
\end{equation}
where 
\begin{equation}
    V_2(m^2_k; Q^2_1, Q^2_2) = \frac{\sqrt{\qty[m^2_K - \qty(\sqrt{Q^2_1} - \sqrt{Q^2_2})^2]\qty[m^2_K - \qty(\sqrt{Q^2_1} + \sqrt{Q^2_2})^2]}}{8\pi m^2_K},
\end{equation}
\begin{equation}
    V_3\qty(Q^2_1)=\frac{Q^2_1}{256\pi^3}\qty[1 - \frac{m^4_{\pi}}{Q^4_1} + 2\frac{m^2_{\pi}}{Q^2_1}\ln{\frac{m^2_{\pi}}{Q^2_1}}],
\end{equation}
and 
\begin{equation}
    V_2(Q^2_2) = \frac{1}{8\pi Q^2_2}\sqrt{Q^2_2\qty(Q^2_2  - 4m^2_{\pi})}.
\end{equation}

The integration in \eqref{1} should be performed in the following domain:
\begin{equation}
    \qty(m_K- \sqrt{Q^2_2})^2 > Q^2_1 > m^2_{\pi}, \; \qty(m_K - m_{\pi})^2 > Q^2_2 > (2  m_{\pi})^2.
\end{equation}

Introducing  dimensionless variables $x = \qty(Q_1/m_K)^2$ and $y = \qty(Q_2/m_K)^2$ we obtain:
\begin{align}
    V_5 = \frac{m^6_K}{2^{16}\pi^7}\int\limits^{\qty(1 - m_{\pi}/m_K)^2}_{\qty(2m_{\pi}/m_K)^2} &dy\int\limits^{\qty(1-\sqrt{y})^2}_{\qty(m_{\pi}/m_K)^2} dx \cdot x\qty[1 - \frac{\qty(m_{\pi/}m_K)^4}{x^2} + 2\frac{\qty(m_{\pi}/m_K)^2}{x}\ln{\frac{\qty(m_{\pi}/m_K)^2}{x}}]\times\\ \nonumber
    & \times \sqrt{1 - 4\frac{\qty(m_{\pi}/m_K)^2}{y}}\sqrt{\qty[1 - \qty(\sqrt{x} - \sqrt{y})^2]\qty[1- \qty(\sqrt{x} + \sqrt{y})^2]}.
\end{align}

Performing the numerical integration we get:

\begin{equation}\label{7}
    V_5 = 2.99 \times 10^{-6}\frac{m^6_K}{2^{16}\pi^7} \equiv \phi \frac{m^6_K}{2^{16}\pi^7}.
\end{equation}
When calculating $V_5$ in \cite{pionium} we took two-particles phase space of $e^+$ and $\nu$ and nonrelativistic phase space of three $\pi^0$-mesons. The result is given by Eq.(8) in \cite{pionium}. It coincides with \eqref{7} within $10\%$.

Now let us take into account a resonance $f_0$ in the system of two $\pi^0$ mesons. In order to do this we multiply the integrand of $V_5$ by the Breit-Wigner factor

\begin{equation}\label{8}
    \frac{\Gamma_i\Gamma_f/4}{\qty(\sqrt{Q^2_2} - M)^2 + \Gamma^2/4},
\end{equation}
where $M$ is a mass of the resonance, $\Gamma$ is its width, $\Gamma_i$ and $\Gamma_f$ are partial $f_0$ widths to initial and final states. Both states are $\pi^0\pi^0$, and $\Gamma_i = \Gamma_f = \Gamma/3$. While final state is $\pi^0\pi^0$, initial state can be a pair of charged pions as well since $f_0$ is isosinglet. So we substituted $\Gamma_i = \Gamma$ and $\Gamma_f = \Gamma/3$. We consider the following mechanism of $f_0$ resonance production: $K^+ \to \pi^0\qty{\pi^0\pi^0 \;\text{or} \; \pi^+\pi^-}e^+ \nu \to \pi^0 f_0 e^+ \nu \to 3\pi^0 e^+ \nu$.
The result of the integration corresponds to the resonance contribution into the width of $K^+ \to 3\pi^0  e^+ \nu$ decay. Let us start with imaginary resonance whose mass equals $300~\text{MeV}$ and make a gedanken experiment varying its width from $0.01~\text{MeV}$ to $1000~\text{MeV}$. Variation of the numerical factor in \eqref{7} is presented in the Table 1\footnote{In these estimates we substituted $\Gamma_i =\Gamma$, $\Gamma_f = \Gamma/3$ which corresponds to the case of $f_0$ meson.}

\begin{table}[H]
\centering
\caption{Dependence of factor $\phi$ on the resonance width $\Gamma$}
\begin{tabular}{ |c|c|c|c|c|c|c| } 
\hline
$\Gamma\text{(MeV)}$ & 0.01 & 0.1 & 1 & 10 & 100 & 1000 \\
\hline
\multirow{1}{2em}{$\phi$} & $2.6\times 10^{-10}$ & $2.6\times 10^{-9}$ & $2.5\times 10^{-8}$ & $2.3\times 10^{-7}$ & $9.0\times 10^{-7}$ & $10\times 10^{-7}$ \\ 
\hline
\end{tabular}
\end{table}
We see that as far as the resonance is narrow ($\Gamma \ll M$), the $K^+$ decay width grows linearly with the resonance width. It could be foreseen since for the narrow resonance the expression \eqref{8} equals $\pi\qty(\Gamma_i\Gamma_f/\Gamma)\delta\qty(\sqrt{Q^2_2} - M)/2$. (The $\delta$-function effectively reduces $V_5$ to $V_4$, as it was done in
\cite{pionium}).

However, when the width of the resonance approaches the value of its mass (the case of $f_0$), the factor \eqref{8} becomes close to one: such a resonance does not enhance the decay probability.
For the contribution of $f_0$ substituting $M=\Gamma=500~\text{MeV}$, $\Gamma_i =\Gamma$, $\Gamma_f = \Gamma/3$
we get $\phi = 5.8\times 10^{-7} $ (compare with \eqref{7}).

In this way we obtain the following "no-go" statement: neither narrow nor wide resonance does enhance the probability of the decay. This statement is valid as long as there is no dominating mechanism of the direct resonance production. In the case of pionium electromagnetic interactions not only form bound states but also change the threshold behaviour of the wave function. This phenomena was investigated in \cite{vainstein} in connection with the positronium states contribution to the muon anomalous magnetic moment. According to \cite{vainstein} the positronium contribution is completely cancelled by the change of the threshold behaviour of the wave function (the so-called Sommerfeld-Gamow-Sakharov factor). It looks like this phenomenon  is universal for the threshold behaviour and is applicable to the case of $f_0$. 

Our conclusion is: the resonance $f_0$ does not enhance the $K^+ \to 3\pi^0  e^+ \nu$ decay probability.

We are grateful to A.E. Bondar for the stimulating questions.

The work is partly supported by the RSCF grant No 22-12-00051-П.


\begin{thebibliography}{}

\bibitem{oka}
A.V.Artamonov et al., (The OKA collaboration). The upper limit on the $K^+ \to \pi^0\pi^0\pi^0 e^+ \nu$ decay. JETP Lett. 120 (2024) 8, 554-558; 
\href{https://doi.org/10.1134/S0021364024603415}{https://doi.org/10.1134/S0021364024603415.}

\bibitem{blaser}
S. Blaser. $K_{e5}$ decays in chiral perturbation theory. Phys.Lett.B.345 (1995) 287. \href{https://doi.org/10.1016/0370-2693(94)01627-O}{https://doi.org/10.1016/0370-2693(94)01627-O.}

\bibitem{pionium}
E.K.Karkaryan, K. V. Kiselev, V. F. Obraztsov, I. V. Surkov, M. I. Vysotsky. On the $K^+ \to \pi^0\pi^0\pi^0 e^+ \nu$ decay. Bull. Lebedev Phys. Inst. 5 (2025) 60. [In Russian] \href{https://arxiv.org/abs/2412.15407#}{arXiv:2412.15407.} 


\bibitem{rpp}
R.~L.~Workman [Particle Data Group], \href{	https://doi.org/10.1093/ptep/ptac097}{PTEP \textbf{2022} (2022), 083C01 doi:10.1093/ptep/ptac097.}

\bibitem{vainstein}
K.~Melnikov, A.~Vainshtein, M.~Voloshin. Remarks on the effect of bound states and threshold in $g-2$. Phys.Rev.D 90 (2014) 1, 017301. \href{https://doi.org/10.1103/PhysRevD.90.017301}{https://doi.org/10.1103/PhysRevD.90.017301.}

\end{thebibliography}
\end{document}